# Modeling of Magnetostriction of Soft Elastomer

**Andriushchenko Petr, Afremov Leonid, Chernova Mariya**

Departament of Theoretical and Experimental Physics, The School of Natural Sciences, Far Eastem Federal University, 8, Sukhanova str., 690950, Vladivostok, Russian Federation

e-mail: pitandmind@gmail.com, afremov.ll@dvfu.ru

**Abstract**. Small magnetic particles placed in a relatively soft polymer (with elastic modulus E ~ 10 ÷ 100 kPa) are magnetically soft elastomers. The external magnetic field acts on each particle which leads to microscopic deformation of the material and consequently to changing of its shape – magnetostriction. For purposes of studying of magnetostriction the model of movable cellular automata (MCA), in which a real heterogeneous material is an ensemble of interacting elements of finite size – automata, is used. It's supposed to be that the motion of each automata can be described by Newton's Second law. The force acting on the i-th automata consists of the following components: volume-dependent force acting on the automata i which is caused by pressure from the surrounding automata; force of an external magnetic field acting on the i-th automata with some magnetic moment; and normal and tangential interaction force between a pair of i and j automata. This approach was used for modeling of magnetostriction elastomer.

## 1. Introduction

At present time there is a necessity for new magnetic materials with properties, substantially depend on a structure (for example, distribution of particles magnetic moments, volumes, coordinates, critical fields, properties of a medium in which particles are placed) and on external influences (magnetic field, pressure, temperature, etc.). MAE can be considered as a class of such "smart"-materials, which one is a composite material, consisting of a flexible polymer medium and the filler - dispersed magnetic powders.

The medium of the MAE is made based on a huge variety of polymeric materials, in particular, it may be polystyrene (PS) [1], polyacrylamide [2], alginate [3,4], polydimethylsiloxane (PDMS) [5,6], ''SIEL'' produced by GNIIChTEOS [6], soft polyurethane (PU), obtained from polyether polyols VORALUX® HF 505 [7] etc.

In the past few years, MAE have attracted growing attention and have been considered for applications as adaptive stiffness elements in vibration absorbers [8], automotive bushing, piezoelectric power actuators [9], as well as in other areas, such as the car industry, medicine (where their properties can be used for cancer therapy by hyperthermia) [10,11], construction, household appliances and etc.

Some methods have been proposed to describe the behavior of magnetorheological (MR) materials. Ginder at al. [12] analyzed average magnetic induction using finite element analysis and computed the shear stresses from the field using Maxwell's stress tensor. A simple dipole model, based on the magnetic interactions between two adjacent particles was extended, to approximate magnetorheological elastomer (MRE) performances [13]. The saturated field-induced shear modulus by considering the interactions in a single particle chain were calculations by Davis [14]. Finite length model based on the microstructural observation of the particles forming the discontinuous and finite length column structures in polymer matrices was presented [15,16].

Most of these models exhibited good accuracy in predicting MR effects caused by embedded ferromagnetic particles, but almost none of these models do not research the fact of the deformation of the sample under the influence of a uniform field. In this regard, the development of a model, formalization and algorithmization of task for modelling of processes, which take place in MAE under an external magnetic field, is necessary.

In this work, an approach to the simulation of MAE is proposed based on a mathematical model of single-domain particles. All calculations were made with particles of the real size. Basic features of the dependence of MAE deformation on a uniform external magnetic field were discussed.

## 2. Method description

To study the magnetostriction of magnetic sponges we use the method of movable cellular automata [17] that combines two following approaches:
- Method of particles, in which, based on the Lagrange formalism, the motion of individual environmental particles is considered;
- Method of cellular automata based on the Euler approach, in which the time-to-time variation of the properties of elements of the fixed uniform grid is studied in relation to the state of surrounding elements.

In the framework of the method of movable cellular automata, the simulated system is represented by an assembly of interacting automata (elements of finite size). The concept of the method is based on the introduction of a new state type – the state of a pair of automata. This allowed to use the spatial variable as a switching parameter. This parameter is represented by $h^{ij}$ - the overlapping of a pair of spherical automata with diameters $d^i$ and $d^j$:

$$h^{ij} = r^{ij} - r_0^{ij}, \qquad (1)$$

where $r_0^{ij} = (d^i + d^j)/2$, $r^{ij}$ is the distance between the centers of the adjacent automata of equilibrium and strain state, respectively. In the simplest case, there are two states of pairs: interconnected $h^{ij} < h_{max}^{ij}$ and unconnected $h^{ij} > h_{max}^{ij}$, where $h_{max}^{ij}$ is a typical condition for this model.

Let's assume that the automata have a spherical shape with a diameter $d_i$ and $N$ automata from the assembly are $N_m$ magnetic. We assume that the magnetic nanoparticle-automata magnetized uniformly. Such magnetic nanoparticle (automata with $i$ number) will be surrounded by $z$ neighbors (see Figure 1). The number of neighbours is determined by the dimension of a problem and packing density. For example, in the case of 2-dimensional problem z = 4 for a simple and z = 6 for a close-packed system shown in Fig. 1.

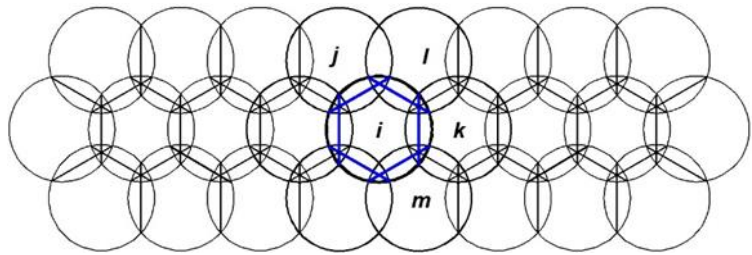

**Figure 1**. Illustration to the model of cellular automata

Let's assume that a simulated by mobile cellular automata environment can be described in terms of classical mechanics. Then, it is justified by the Newton's second law and the law of dynamics of rotational motion:

$$m_i \frac{d^2 \boldsymbol{r}_i}{dt^2} = \boldsymbol{F}_i = \boldsymbol{F}_i^{\Omega} + \sum_{j=1}^{z}\left(\boldsymbol{F}_{ij}^{\text{norm}} + \boldsymbol{F}_{ij}^{\text{tang}}\right) + \boldsymbol{F}_i^{\text{m}}, \tag{2}$$

$$J_i \frac{d^2 \boldsymbol{\theta}_i}{dt^2} = \boldsymbol{K}_i = \sum_{j=1}^{z}\left[\mathbf{n}_{ij}, \boldsymbol{F}_{ij}^{\text{tang}}\right] q_{ij} S_{ij} + [\boldsymbol{r}_i, \boldsymbol{F}_i^{\text{m}}], \tag{3}$$

where $\boldsymbol{r}_i$ here is the radius vector of the center of the mass of the automata $i$, $\boldsymbol{F}_i^{\Omega} = \sum_{j=1}^{z} P_j S_{ij} \mathbf{n}_{ij}$ - the volume-dependent force effecting on the automata $i$ and provided by the pressure from surrounding automata, $P_j = -K_j \frac{\Omega_j - \Omega_j^0}{\Omega_j^0} = -K_j D \bar{\varepsilon}_j$ - the pressure of the $j$-th neighboring automata, $\Omega_j^0$- is the initial equilibrium automata volume, $\Omega_j$ - the current automata volume, $K_j$ - the module of uniform compression of the material of this automata, $D$ - the coefficient depending on the dimension of the problem ($D = 1, 2, 3$), $\bar{\varepsilon}_j = \sum_{j=1}^{z} \varepsilon_{ij}/z$ - the average strain, $\varepsilon_{ij} = (2q_{ij} - d_i)/d_i$, where $S_{ij}$- the contact area of the $i$-th automata with the $j$-th and $q_{ij}$ - the distance from the center of the $i$-th automata to the surface of the $j$-th; $\mathbf{n}_{ij} = (\boldsymbol{r}_i - \boldsymbol{r}_j)/r^{ij}$ - the normal vector between the $i$-th and $j$-th automata, z – the number of nearest neighbors. $\boldsymbol{F}_{ji}^{\text{norm}}$ - pair force of normal interaction between $i$ and $j$ automata, which in approximation of the Hooke's law is determined by a superposition of pair forces of elastic (first component) and viscous interaction:

$$\boldsymbol{F}_{ij}^{\text{norm}} = -\left(2G(\varepsilon_{ij} - \bar{\varepsilon}_i) + \eta \frac{(\boldsymbol{v}_j - \boldsymbol{v}_i, \mathbf{n}_{ij})}{r^{ij}}\right) S_{ij} \mathbf{n}_{ij}, \tag{4}$$

where $G$ - is the shear modulus, $\eta$ - the viscosity coefficient, $\boldsymbol{v}_i$- the speed of the the $i$-th automata. $\boldsymbol{F}_{ij}^{\text{tang}}$ - force of pair tangential interaction, as well as pair force of normal interaction, elastic shear and viscous components:

$$\boldsymbol{F}_{ij}^{\text{tang}} = -2G[\boldsymbol{\gamma}_{ij} + \boldsymbol{\gamma}_{ji}, \mathbf{n}_{ij}] - \eta \frac{\left[(\boldsymbol{\omega}_{ij} r^{ij} - \boldsymbol{\omega}_i q_{ij} - \boldsymbol{\omega}_j q_{ji}), \mathbf{n}_{ij}\right]}{r^{ij}}, \tag{5}$$

where $\boldsymbol{\gamma}_{ij} = \sum_k [\mathbf{n}_{ij}^{k-1}, \mathbf{n}_{ij}^k] - \boldsymbol{\theta}_i$, $\boldsymbol{\gamma}_{ji} = \sum_k [\mathbf{n}_{ji}^{k-1}, \mathbf{n}_{ji}^k] - \boldsymbol{\theta}_j$ is a full shift angle of the $i$-th and $j$-th automata correspondently, $\boldsymbol{\theta}_i$ and $\boldsymbol{\theta}_j$ - rotation angle of each of the automata, $\boldsymbol{\omega}_{ij} = [\mathbf{n}_{ij}, (\boldsymbol{v}_j - \boldsymbol{v}_i)]/r^{ij}$ and $\boldsymbol{\omega}_i$ - angular speeds of rotation of a pair and the $i$-th automata, respectively, $m_i$ and $J_i$- the mass and moment of inertia of the $i$-th automata, respectively.

External magnetic fields of uniformly magnetized (single-domain) spherical particles–automata are dipole. Therefore, replacing the particle by a point dipole, we assume that the $i$-th particle is exposed by the local field $\boldsymbol{H}(\boldsymbol{r}_i)$ defined as the sum of the external magnetic field $\boldsymbol{h}$ and fields of magnetic dipole interaction caused by other particles:

$$\boldsymbol{H}(\boldsymbol{r}_i) = \boldsymbol{h} + \sum_j \left(-\frac{\boldsymbol{\mu}^j}{r_{ij}^3} + \frac{3(\boldsymbol{\mu}^j, \boldsymbol{r}_{ij})\boldsymbol{r}_{ij}}{r_{ij}^5}\right), \boldsymbol{r}_{ij} = \boldsymbol{r}_i - \boldsymbol{r}_j. \tag{6}$$

$\boldsymbol{\mu}^j$ here is the magnetic moment of the given $j$-th particle. Since the energy of the point magnetic dipole is $V^i(\boldsymbol{r}_i) = -\left(\boldsymbol{\mu}^i, \boldsymbol{H}^i(\boldsymbol{r}_i)\right)$, due to the homogeneity of the magnetic moment $\boldsymbol{\mu}^j$ the magnetic force on the $i$-th particle–automata is:

$$\boldsymbol{F}_i^m = -\boldsymbol{\nabla} V^i(\boldsymbol{r}_i) = (\boldsymbol{\mu}^i, \boldsymbol{\nabla})\boldsymbol{H}^i(\boldsymbol{r}_i). \tag{6}$$

The external field $\boldsymbol{h}$ mediately affects magnetic force. It leads to a change in magnetic moments of the particles–automata, which leads to a complete change of the magnetic interaction between magnetic moments of the particles and, as a result, to a change of the force $\boldsymbol{F}_i^m$.

## 3. Simulation algorithm

The main part of the simulation algorithm in this model is the determination of the positions and velocities of the automata of each $k$-th iteration:

$$\boldsymbol{r}_i(t^{k+1}) = \boldsymbol{r}_i(t^k) + \boldsymbol{V}_i(t^k)\Delta t, \tag{7}$$
$$\boldsymbol{V}_i(t^{k+1}) = \boldsymbol{V}_i(t^k) + \boldsymbol{F}_i/m_i, \tag{8}$$
$$\boldsymbol{\theta}_i(t^{k+1}) = \boldsymbol{\theta}_i(t^k) + \boldsymbol{\omega}_i(t^k)\Delta t, \tag{9}$$
$$\boldsymbol{\omega}_i(t^{k+1}) = \boldsymbol{\omega}_i(t^k) + \boldsymbol{K}_i/J_i, \tag{10}$$

where $t^{k+1} = t^k + \Delta t$, $\Delta t$ - the time of one iteration commensurate with the time of passage of a sound through the automata: $\tau = d_i/C$ ($C = \sqrt{G/\rho}$ - the speed of environmental sound, $\rho$ - the density of the environment).

Figure 2 illustrates an example of the media magnetostriction. The proposed approach can be used for simulating the magnetostriction of soft elastomer.

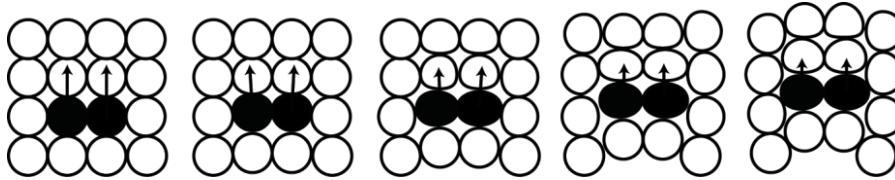

**Figure 2.** Examples of pre-simulation. The magnetic particles are marked black, medium are marked white.

## 4. Acknowledgements

The work is supported by grant of Ministry of Education and Science 02.740.11.0549 (Reference number 2010-1.2.2-214-005-006).